\documentclass{ifacconf}

\usepackage{natbib}        

\usepackage[utf8]{inputenc}
\usepackage[T1]{fontenc}
\usepackage{amsmath}
\usepackage{amsfonts}
\usepackage{amssymb}
\usepackage{graphicx}
\usepackage{physics}
\usepackage{commath}
\usepackage{xcolor}
\usepackage{float}
\usepackage{optidef}
\usepackage[shortlabels]{enumitem}
\usepackage{algorithm}
\usepackage{algpseudocode}
\usepackage{subfig}
\usepackage{import}

\usepackage{mathtools}
\DeclarePairedDelimiterX{\inp}[2]{\langle}{\rangle}{#1, #2}
\DeclareMathOperator{\vect}{vec}

\newtheorem{proposition}{Proposition}
\newtheorem{assumption}{Assumption}

\newtheorem{remark}{Remark}

\usepackage{tikz}
\usepackage{pgfplots}
\pgfplotsset{compat=1.6}
\usepgfplotslibrary{groupplots} 
\usepgfplotslibrary{colorbrewer}
\pgfplotsset{every axis/.append style={cycle list/Dark2}, ylabel near ticks, xlabel near ticks}
\pgfplotsset{every axis plot/.append style={thick, line join=round}}
\pgfplotsset{every axis legend/.append style={legend cell align=left,align=left, font=\small, fill opacity=0.7,draw opacity=1,text opacity=1}}
\pgfplotsset{every axis title/.append style={font=\itshape}}
\pgfkeys{/pgf/number format/.cd,1000 sep={}} 
\pgfplotsset{every axis/.append style={yticklabel style={/pgf/number format/fixed,/pgf/number format/precision=5},scaled y ticks=false,}} 
\pgfplotsset{every axis/.append style={xticklabel style={/pgf/number format/fixed,/pgf/number format/precision=5},scaled x ticks=false,}} 

\usetikzlibrary{external}
\DeclareMathOperator*{\argmin}{\arg\min}

\begin{document}
\begin{frontmatter}
	
	\title{Measurement-Based Control for Minimizing Energy Functions in Quantum Systems}

	
	
	\author[AAU]{Henrik Glavind Clausen} 
	\author[AAU]{Salahuddin Abdul Rahman} 
	\author[Kadir]{Özkan Karabacak}
	\author[AAU]{Rafal Wisniewski}
	
	\address[AAU]{Section for Automation and Control, Department of Electronic Systems, Aalborg University, Aalborg, Denmark (e-mail: \{hgcl, saabra, raf\}@es.aau.dk).}
	\address[Kadir]{Department of Mechatronics Engineering, Kadir Has University, Istanbul, Turkey (e-mail: ozkan.karabacak@khas.edu.tr).}
	
	\begin{abstract}                
		In variational quantum algorithms (VQAs), the most common objective is to find the minimum energy eigenstate of a given energy Hamiltonian. In this paper, we consider the general problem of finding a sufficient control Hamiltonian structure that, under a given feedback control law, ensures convergence to the minimum energy eigenstate of a given energy function. By including quantum non-demolition (QND) measurements in the loop, convergence to a pure state can be ensured from an arbitrary mixed initial state. Based on existing results on strict control Lyapunov functions, we formulate a semidefinite optimization problem, whose solution defines a non-unique control Hamiltonian, which is sufficient to ensure almost sure convergence to the minimum energy eigenstate under the given feedback law and the action of QND measurements. A numerical example is provided to showcase the proposed methodology.
	\end{abstract}
	
	
\end{frontmatter}

\section{Introduction}
In variational quantum algorithms (VQAs), the objective is to optimize a parameterized quantum circuit with respect to some cost function (see, e.g., \cite{cerezo2021} for a recent survey of VQAs). Two important examples include the variational quantum eigensolver (VQE) and the quantum approximate optimization algorithm (QAOA), in which the goal is to find the minimum energy eigenstate of a given Hamiltonian. If this Hamiltonian encodes the solutions to a combinatorial optimization problem, then the QAOA can be used to solve this combinatorial optimization problem, making the QAOA of broad interest within many branches of, e.g., the technical sciences and economy. 

VQAs are hybrid algorithms, where part of the algorithm (the quantum circuit) is run on the quantum computer, and another part (the optimization of the circuit parameters) is run on a classical computer. One of the bottlenecks in the VQA framework is the classical optimizer, which is needed to update the circuit parameters. Generally, this problem is highly non-convex, making it difficult to find optimal parameters. For a recent control-inspired perspective on VQAs, see, e.g., \cite{magann2021a,magann2022}. In particular, \cite{magann2022} propose using ideas from Lyapunov control as replacement for the classical optimization routine, by viewing the circuit parameters as the control signal of a discrete-time quantum dynamical system. This approach is, however, only ensured to converge to the optimal solution under some harsh assumptions known from quantum control theory.

Lyapunov-based control for quantum systems has been studied extensively over the last two decades. For pure quantum systems governed by the Schrödinger equation, \cite{grivopoulos2003} provide several results detailing sufficient assumptions to ensure convergence to a target pure state for a Lyapunov function quadratic in the state. Likewise, \cite{mirrahimi2005} provide similar results for convergence to a target trajectory based on a Lyapunov function defined in terms of the state error. In continuation of this, \cite{kuang2008} provide a unifying view of different choices of Lyapunov functions. In \cite{beauchard2007}, an implicit Lyapunov scheme is used to control systems that are locally un-controllable. These results are extended to mixed states by \cite{altafini2007} and \cite{wang2010a,wang2010b}. Here, it is shown that under the same assumptions as for the pure-state case, convergence to the (unique) state that minimizes a given Lyapunov function cannot be ensured from arbitrary mixed states (for a recent overview, see, e.g., Chapter 8 in \citep{dalessandro2021}). As will be elaborated on shortly, we show, as part of the contribution of the present work, that this issue can be alleviated by introducing quantum measurements in the loop. 

As projective von Neumann measurements collapse the quantum state, closed-loop control may seem difficult to achieve. However, by entangling the system of interest to an ancillary probe system and subsequently performing a projective measurement on just the probe system, one can achieve an indirect or generalized form of measurement that results in a less-severe back-action on the primary system. One particular form of this generalized measurement is known as a \emph{quantum non-demolition} (QND) measurement, which is characterized by commuting with a particular observable/operator of interest \citep{amini2012}. Repeatedly applying QND measurements then induces a stochastic process on the space of quantum states. By means of a quantum filter, it is possible, based on the QND measurement outcomes, to estimate the quantum state on-line. The use of QND measurements thus paves the way for actual measurement-based state-feedback control of quantum systems. There is a vast literature on continuous-time QND measurements (see, e.g., \cite{belavkin1999}), but in this paper we limit ourselves to the discrete-time case. In the present work, our main motivation for including QND measurements is the property that they will purify the state along the trajectory, making it possible to converge to a pure state from any mixed state.

Feedback control with discrete-time QND measurements was investigated by \cite{dotsenko2009} and \cite{mirrahimi2009} with a particular experimental setup, commonly referred to as the photon box experiment, as case study, where QND measurements are used in combination with Lyapunov-based control to stabilize a desired pure quantum state. These results were later extended and generalized by \cite{amini2011,amini2012,amini2013} to account for delays in the system and to provide concrete design rules for the Lyapunov function to ensure convergence.

In all of the above studies, the control problem has been posed as designing a control law that ensures convergence to a known target state, which, in the Lyapunov-based case, mostly reduces to designing a suitable Lyapunov function. In this paper, however, we reverse this design philosophy to fit the VQA framework and pose the following problem: \emph{Given a fixed energy (Lyapunov) function, can we design a quantum system that ensures convergence (from any initial state) to the state which minimizes this energy function?} This problem statement, in the case of VQE and QAOA, is equivalent to ask: \emph{Can we design a quantum circuit and a feedback control law (i.e., circuit parameter update rule) that ensures convergence to the minimum energy eigenstate of a given Hamiltonian?}

The contribution of the present work is therefore as follows: In the present work, we investigate whether the use of discrete-time QND measurements can improve the convergence properties when aiming to stabilize the system in the state which minimizes a given energy function. In particular, our main contribution is the formulation of a semidefinite program (SDP), whose solution will be used to construct a sufficient control Hamiltonian (or, equivalently, parameterized quantum circuit in the VQA setting) that ensures (based on the results from \cite{amini2013}) that the measurement-induced process with a particular control choice will converge to the minimizer of an arbitrary but known energy function. These results thus pave the way towards better design choices of quantum circuits in VQAs.

The rest of the paper is organized as follows. In Section \ref{sec:deterministic}, we establish our definitions and notational conventions, and recall the main results from existing literature on Lyapunov-based quantum control in the measurement-free case as well as in the QND measurement case. In Section \ref{sec:sdp}, we provide our main result in the form of an SDP, whose solution will be used to construct a control Hamiltonian for the problem at hand. In Section \ref{sec:example}, we provide a numerical example, and, lastly, in Section \ref{sec:discussion} we discuss perspectives and future work.

\section{Background}
\label{sec:deterministic}
Denote the state space $\mathcal{H}=\mathbb{C}^N$ with orthonormal basis $\mathcal{F} = \{\ket{n}\}_{n\in\{0,\dots,n_\text{max}\}}$ and let the space of density operators be denoted by $\mathcal{X}=\{\rho \in \mathbb{C}^{N\times N} \mid \rho \geq 0, \Tr(\rho) = 1\}$ with $N=n_\text{max}+1$. We will in the following represent all operators in the $\mathcal{F}$ basis.

In the following, we first present some basic results from closed quantum systems without measurements, followed by a discussion on results for quantum systems driven by QND measurements.

\subsection{Closed quantum systems without measurements}
In the continuous-time setting, the control input $u(t)$ enters the dynamics through the Schrödinger equation
\begin{align}
	\dot{\psi}(t) = -i(H_0 + u(t) H_1)\psi(t), \quad \psi(0) = \psi_0,
\end{align}
or, given in the density formalism, 
\begin{align}
	\dot{\rho}(t) = -i[H_0+u(t)H_1,\rho(t)], \quad \rho(0)=\rho_0,
\end{align}
where $[A,B]:=AB-BA$ is the commutator of operators $A$ and $B$. Here, $H_0$ is the drift Hamiltonian, i.e., the Hamiltonian governing the unforced dynamical system, and $H_1$ is the control Hamiltonian. Both are assumed to be time-invariant. We will throughout the paper assume $H_0$ and $H_1$ to be non-commuting, i.e., $[H_0,H_1]\neq 0$.

Let $P$ denote the operator of which we want to minimize the energy and let $V(\rho) = \Tr(P\rho)$ be our Lyapunov function. In the setting of VQAs, this operator $P$ can, for instance, represent the energy levels of a molecule for which we want to find the ground state (as in the VQE), or it could represent the possible solutions to a combinatorial optimization problem (as in the QAOA). We now introduce our first assumption.
\begin{assumption}\label{assumption:1}
	The matrices $P$ and $H_0$ are diagonal in the basis $\mathcal{F}$ with eigenvalues $p_0,\dots,p_{n_{\text{max}}}$ and $h_0,\dots,h_{n_{\text{max}}}$, respectively.
\end{assumption}
The problem is then to design a state-feedback law $u(\rho)$ that stabilizes the state $\rho$ at $\rho^\star = \arg\min_{\rho\in\mathcal{X}} \Tr(P\rho)$.

It is easily shown that the derivative of $V(\rho)$ along trajectories is given by
\begin{align}
	\nabla V \cdot \dot{\rho} = -i\Tr([P,H_1]\rho(t))u(t).
\end{align}
Thus, choosing the control law as
\begin{align}
	u(t) = i \kappa \Tr([P,H_1]\rho(t))
\end{align}
for $\kappa > 0$ will ensure that the Lyapunov function is non-increasing along trajectories, i.e., $\nabla V \cdot \dot{\rho} \leq 0$. By LaSalle's invariance principle, the state $\rho(t)$ then converges to the largest invariant subset of $\{\rho\in\mathcal{X} \mid \nabla V \cdot \dot{\rho} = 0\}$.

It is well-known (see, e.g., \cite{grivopoulos2003,altafini2007,wang2010a,wang2010b,dalessandro2021}) that the following three assumptions, together with Assumption \ref{assumption:1}, are sufficient to ensure that the largest invariant set is exactly given by density matrices diagonal in the $\mathcal{F}$ basis. 
These assumptions are as follows:


\begin{assumption}\label{assumption:1b}
	The spectrum of $P$ is non-degenerate, i.e., $p_i \neq p_j$ for all $i \neq j$.
\end{assumption}

\begin{assumption}\label{assumption:2}
	The drift Hamiltonian $H_0$ is strongly regular, i.e., $(\lambda_i - \lambda_j) \neq (\lambda_k - \lambda_l)$ for all $(i,j)\neq (k,l)$ with $\lambda_i$ denoting the $i$'th eigenvalue of $H_0$. In other words, the eigenvalue gaps of the drift Hamiltonian $H_0$ are distinct.
\end{assumption}

\begin{assumption}\label{assumption:3}
	The control Hamiltonian $H_1$ is fully connected, meaning all off-diagonal entries of $H_1$ are non-zero, i.e., $H_{1}^{i,j} \neq 0$ for all $i\neq j$.
\end{assumption}

In this paper, we will, however, consider discrete-time systems of the form
\begin{align}
	\rho_{k+1} = U(u_k) \rho_k U(u_k)^\dagger,
\end{align}
where $k\in\mathbb{N}$ is the (discrete) time index and $U(u)$ is a unitary operator parameterized in terms of control signal $u$. In particular, we consider $U(u)$ of the form
\begin{align}
	U(u) = \mathrm{e}^{-iH_0} \mathrm{e}^{-iH_1u},
\end{align}
where $H_0$ and $H_1$ denote the drift and control Hamiltonian, respectively, similarly to the continuous-time case. This model can be seen as the result of applying two unitary gates $U_0=\mathrm{e}^{-iH_0}$ and $U_1(u) =\mathrm{e}^{-iH_1 u}$ in a quantum circuit, similarly to a QAOA circuit, or it can be seen as a (rough) piece-wise constant discretization of a continuous-time system with unit discretization time. In our work, we will stick to the first interpretation to avoid making assumptions on when the time discretization is valid. For an interpretation of the latter case, we refer to \cite{magann2022}. 

Discrete-time systems of this form are rarely studied in the literature (see, e.g., \cite{magann2021a}), so in the following we will first establish that Assumptions \ref{assumption:1}--\ref{assumption:3}, together with a feedback law similar to \eqref{eq:fblaw}, are in fact also sufficient to ensure convergence to $\rho^\star$ in the discrete-time case.

We will make use of the same Lyapunov function as above. In discrete-time, we want to ensure that $V(\rho)$ is decreasing between time steps, i.e., $V(\rho_{k+1}) - V(\rho_k) \leq 0$.

By using a first-order approximation of the exponential terms, i.e.,
\begin{align}
	\mathrm{e}^{-iH_0} \mathrm{e}^{-iH_1u} = \mathrm{e}^{-iH_0} (I -i H_1 u) + O(u^2),
\end{align}
we get
\begin{equation}
	\begin{aligned}
		&V(\rho_{k+1}) - V(\rho_k)\\
		=& \Tr(P \mathrm{e}^{-iH_0} \mathrm{e}^{-iH_1u_k} \rho_k \mathrm{e}^{-iH_0} \mathrm{e}^{-iH_1u_k}) - \Tr(P\rho_k) \\
		=& \Tr(P\mathrm{e}^{-iH_0}(\rho_k +i[\rho_k,H_1 u_k])\mathrm{e}^{iH_0}) - \Tr(P\rho_k) + O(u_k^2)\\
		=& -i \Tr([P,H_1]\rho_k) u_k + O(u_k^2),\label{eq:lyapunov_discrete}
	\end{aligned}
\end{equation}
where we in the last equality used the fact that $P$ and $H_0$ commute and the relation $\Tr(P[\rho,H_1])=-\Tr([P,H_1]\rho)$.

As in the continuous-time case, we can pick the control law 
\begin{align}
	u_k = i \kappa \Tr([P,H_1]\rho_k) \label{eq:fblaw}
\end{align}
with $\kappa>0$ to ensure that $V(\rho_{k+1}) - V(\rho_k) \leq 0$. For the first-order approximation to be valid, $\kappa$ should be sufficiently small. A rigorous analysis of this is out of scope for the current work, but we refer to \cite{magann2022} for a comparable analysis in terms of discretization time.

It thus remains to characterize the largest invariant subset of $\mathcal{W}=\{\rho \in \mathcal{X} \mid V(\rho_{k+1}) - V(\rho_k) = 0 \}$. For any $\rho\in\mathcal{W}$ being a stationary point of $V(\rho)$, we have from \eqref{eq:lyapunov_discrete} and \eqref{eq:fblaw} that the control $u$ must be zero. Thus, for $\mathcal{V}\subseteq\mathcal{W}$ to be an invariant subset of $\mathcal{W}$, we must have that any $\bar{\rho}\in\mathcal{V}$ must stay (for all time $k$) in $\mathcal{V}$ under the action of the unforced dynamics, i.e., $U(0)^k\bar{\rho}\, U(0)^{\dagger k} \in \mathcal{V}$ for all $k\in\mathbb{N}$. 
For $U(0)=\mathrm{e}^{-iH_0}$ we thus have the condition
\begin{align}
	-i \kappa \Tr([P,H_1]\bar{\rho}) = -i \kappa \Tr([P,H_1]\mathrm{e}^{-iH_0k} \bar{\rho}\, \mathrm{e}^{iH_0k}) = 0.
\end{align}
Using the fact that $P$ and $H_0$ are diagonal with diagonal elements $p_0,\dots,p_{n_\text{max}}$ and $h_0,\dots,h_{n_\text{max}}$, respectively, we have
\begin{equation}
	\begin{aligned}
		0 &= \Tr([P,H_1]\mathrm{e}^{-iH_0k} \bar{\rho}\, \mathrm{e}^{iH_0k}) \\
		&= \Tr\Bigg(\sum_{i,j,n} p_n \mathrm{e}^{i(h_j-h_i)k} \bar{\rho}_{i,j} \op{n} H_1 \op{i}{j} \\&\phantom{fuckits}- p_n \mathrm{e}^{i(h_j-h_i)k} \bar{\rho}_{i,j}  H_1 \op{n} \op{i}{j}\Bigg) \\
		&=\Tr(\sum_{i,j} (p_j-p_i) \mathrm{e}^{i(h_j-h_i)k} \bar{\rho}_{i,j} H_1 \op{i}{j}) \\
		&= \sum_{i,j} (p_j-p_i) \mathrm{e}^{i(h_j-h_i)k} \bar{\rho}_{i,j} \Tr(H_1 \op{i}{j}) \\
		&= \sum_{i,j} (p_j-p_i) \mathrm{e}^{i(h_j-h_i)k} \bar{\rho}_{i,j} \mel{j}{H_1}{i}.
	\end{aligned}\label{eq:invariant_set}
\end{equation}
As $\bar{\rho}$ and $H_1$ are Hermitian, we can reduce the above to
\begin{align}
	\Re\left(\sum_{i < j} (p_j-p_i) \mathrm{e}^{i(h_j-h_i)k} \bar{\rho}_{i,j} \mel{j}{H_1}{i}\right) = 0.
\end{align}
From here, it is clear that if
\begin{itemize}
	\item $p_i \neq p_j$ for all $i< j$ (Assumption \ref{assumption:1b}),
	\item $(h_j-h_i)\neq(h_k - h_l) \pmod{2\pi}$ for all $(i,j)\neq (k,l)$ (Assumption \ref{assumption:2}, but modulo $2\pi$),
	\item and $\mel{j}{H_1}{i}\neq 0$ for all $i \neq j$ (Assumption \ref{assumption:3}),
\end{itemize}
then the only solution is the trivial one, i.e., $\bar{\rho}_{i,j}=0$ for all $i<j$, which is equivalent to $\bar{\rho}$ being diagonal. We remark that Assumption \ref{assumption:2} ensures linear independence of the complex exponential functions $f_{i,j}(t):=\mathrm{e}^{i(h_j-h_i)t}$ for $t\in\mathbb{R}$, but for discrete time $k\in\mathbb{N}$, we need to adjust the assumption to be stated in modulo $2\pi$ due to the periodicity of the exponential function and the discrete time index.


Note that, as the evolution of the state is unitary, the spectrum of $\rho_k$ is preserved, meaning that $\rho_k$ and $\rho_0$ are isospectral for any $k$. This implies that if $\rho_0$ is a pure state, then $\rho_k$ is also pure, and if $\rho_0$ is a mixed state, then $\rho_k$ is also mixed.

 
Hence, to conclude, under Assumptions \ref{assumption:1}--\ref{assumption:3}, the largest invariant subset of $\{\rho \in \mathcal{X} \mid V(\rho_{k+1}) - V(\rho_k) = 0 \}$ is the set of all diagonal $\rho$ (or, equivalently, those $\rho$ that commute with $H_0$ and $P$). However, under unitary evolution, the only reachable subset is the set of $\rho$ isospectral to $\rho_0$. Thus, for a fixed $\rho_0$, the state $\rho_k$ will under feedback law \eqref{eq:fblaw} converge to the set 
\begin{align}
	\{\rho \in \mathcal{X} \mid [\rho,P]=0,\quad \mathrm{spec}(\rho) = \mathrm{spec}(\rho_0) \}.\label{eq:inv_set_discrete}
\end{align}

\begin{remark}
	If Assumption \ref{assumption:1b} is not satisfied, i.e., that the spectrum of $P$ is degenerate, the limit set is still given as \eqref{eq:inv_set_discrete}, but the condition $[\rho,P]=0$ no longer implies that $\rho$ is diagonal. In particular, the invariant set is now extended to contain the states that are spanned by the eigenvectors of $P$ corresponding to the degenerate eigenvalues. However, if the minimum eigenvalue of $P$ is unique, and the initial state $\rho_0$ is not contained in the invariant set, then, because $V(\rho)$ is decreasing, the state will eventually converge to the minimum state.
\end{remark} 
When $\rho_0$ is a pure state, the invariant set contains exactly the $N$ eigenstates of $P$. Furthermore, with the spectrum of $P$ being non-degenerate, the Lyapunov function $V(\rho) = \Tr(P\rho)$ has $N$ critical points consisting of one maximum, one minimum and $N-2$ saddle points, located at the $N$ eigenstates of $P$. Thus, when $\rho_0$ is pure, $V(\rho_{k+1})-V(\rho_k)$ is strictly decreasing everywhere except on these critical points, and as saddle points are only stable from a set of initial conditions with zero measure, convergence to $\rho^\star = \argmin_\rho V(\rho)$ is ensured almost everywhere, i.e., $\rho_k$ will converge to $\rho^\star$ from every pure initial state $\rho_0$ except a set of zero measure.

When $\rho_0$ is a mixed state, the situation is more complicated. Because the spectrum of $\rho_0$ is preserved, it is, first of all, not possible to reach the pure state $\rho^\star$ that minimizes $V(\rho)$ if $\rho_0$ is not pure. For generic states, i.e., states with a non-degenerate spectrum, convergence to $\rho_\text{mixed}^\star = \argmin_{\rho\in\{\rho\in\mathcal{X} \mid \mathrm{spec}(\rho) = \mathrm{spec}(\rho_0) \}} V(\rho)$ is, however, still ensured, but for general mixed states with degenerate spectra, this is not the case. For further details, we refer to \cite{wang2010a,wang2010b}.

Thus, to summarize, the feedback law \eqref{eq:fblaw} cannot ensure convergence to $\rho^\star$ in the general mixed state case. In \cite{magann2022}, the authors use the feedback law \eqref{eq:fblaw} to determine parameters in the QAOA algorithm with some success, although they suffer from not adhering to the hard assumptions on $H_1$ (Assumption \ref{assumption:3}). In the following, we show how we can alleviate this problem by introducing quantum non-demolition (QND) measurements in the loop. The main motivation for the use of QND measurements is the property that they will purify the state along the trajectory, making it possible to converge to a pure state from any mixed state. In addition to the advantageous properties in terms of convergence, the QND measurements have the additional benefit of providing information about the current state, which, together with a simple quantum filter, provides a state estimate that can be used directly in the state-feedback control. 

\subsection{Quantum systems with QND measurements}\label{sec:qnd}

Let a generalized quantum measurement $\mathcal{M}$ be constituted by measurement operators $\{M_\mu\}_{\mu\in\{1,\dots,m\}}$ adhering to the completeness relation
\begin{align*}
	I =\sum_\mu M_\mu^\dagger M_\mu. 
\end{align*}

Let the post-measurement state after observing outcome $\mu$ be denoted by $\mathbb{M}_\mu(\rho) = \frac{M_\mu \rho M_\mu^\dagger}{\Tr(M_\mu^\dagger M_\mu \rho)}$, where outcome $\mu$ occurs with probability $p_\mu = \Tr(M_\mu^\dagger M_\mu \rho)$. The measurement $\mathcal{M}$ thus induces a Markov process 
\begin{align}
	\rho_{k+1} = \mathbb{M}_{\mu_k}(\rho_k)\label{eq:qnd_openloop}
\end{align}
on the space of density operators with initial state $\rho_0\in\mathcal{X}$.

Quantum non-demolition (QND) measurements are a special case of the generalized measurements, where all measurement operators $M_\mu$ are simultaneously diagonalizable. In particular, we will assume that they are diagonal in our preferred basis $\mathcal{F}$, i.e., 
\begin{align}
	M_\mu = \sum_{n\in\mathcal{F}} c_{\mu,n} \op{n},\quad \sum_\mu |c|_{\mu,n}^2 = 1\text{ for all }n\in\mathcal{F}.\label{eq:qnd_cmu}
\end{align}
%
QND measurements have several interesting properties. Because the measurement operators are diagonal in the $\mathcal{F}$-basis, they commute with many important operators we have already introduced, namely, $P$ and $H_0$. Because of this property, the Hilbert-Schmidt inner product (trace) between any diagonal operator $A$ and a QND-induced Markov process $\{\rho_k\}$ is a martingale, i.e.,
\begin{align}
	\mathbb{E}[\Tr(A\rho_{k+1}) \mid \rho_k] - \Tr(A\rho_{k}) = 0.
\end{align}
Furthermore, as has also been proved in several places (see, e.g., \cite{amini2011,amini2012}), the process \eqref{eq:qnd_openloop} converges almost surely to an eigenstate $\{\op{n}\}_{n\in\mathcal{F}}$ of $\{M_\mu\}$ with the probability of converging to state $\op{\bar{n}}$ given by $\Pr[n=\bar{n}] = \Tr(\op{\bar{n}}\rho_0)$.

When introducing QND measurements, we will consider the controlled Markov process
\begin{align}
	\rho_{k+1} = U(u_k)\mathbb{M}_{\mu_k}(\rho_k)U(u_k)^\dagger.\label{eq:qnd_model}
\end{align}
%
As in Section \ref{sec:deterministic}, the unitary operator $U(u)$ could have several forms. In the following, we will stick to the form $U(u)=\mathrm{e}^{-iH_1 u}$. Note that we have disregarded the drift Hamiltonian $H_0$ here. Recall that, in the measurement-free case, the main contribution of $H_0$ is in the characterization of the largest invariant set, where it ensures that the invariant set only contain states (mixed or pure) that commute with $H_0$. With QND measurements included in the process, this is no longer needed, as the QND measurements will serve the same purpose, except that they limit the invariant set to contain \emph{only} pure states that commute with $\{M_\mu\}$, meaning mixed states are no longer invariant. 

For the model \eqref{eq:qnd_model}, the feedback law \eqref{eq:fblaw}, together with the assumptions \ref{assumption:1}--\ref{assumption:3} from Section \ref{sec:deterministic}, are no longer sufficient to ensure convergence to $\rho^\star = \argmin_\rho V(\rho)$. In the measurement-free case, the fact that all the non-minimizing eigenstates are saddle points of $V(\rho)$ ensure that the system converges to the minimum. However, because the QND measurements by themselves will drive the state towards a (random) eigenstate of $P$, the previous argument on saddle points no longer hold.

In the following, we will discard Assumptions \ref{assumption:2} and \ref{assumption:3}, and instead introduce the following standard assumption (see, e.g., \cite{amini2011,amini2013}) on the QND measurement operators, which may be seen as the counterpart to Assumption \ref{assumption:1b} on the drift Hamiltonian.
\begin{assumption}\label{assumption:qnd}
	For all $n_1 \neq n_2$, there exists a $\mu$ such that $\abs{c_{\mu,n_1}}^2 \neq \abs{c_{\mu,n_2}}^2$ (with $c_{\mu,n}$ as defined in \eqref{eq:qnd_cmu}).
\end{assumption}
This assumption ensures that the measurement statistics are unique for each eigenstate of $\{M_\mu\}$, which in turn is sufficient to ensure that the only invariant states under $\mathbb{M}_\mu$ $\forall \mu$ are the pure eigenstates themselves (see, e.g., \cite[Theorem~3.1]{amini2011} for the proof).

In \cite{amini2011,amini2013}, the authors propose the Lyapunov function
\begin{align}
	V_\varepsilon(\rho) = \Tr(\rho P) -\frac{\varepsilon}{2} \sum_n (\!\ev{\rho}{n})^2 
\end{align}
and the feedback law
\begin{align}
	u_k = \argmin_{u\in[-\bar{u},\bar{u}]} \mathbb{E}\left[V_\varepsilon(\rho_{k+1}) \mid \rho_k, u\right],\label{eq:fblaw_amini2013}
\end{align}
where $\varepsilon>0$ ensures that $V_\varepsilon(\rho)\eval_{u=0}$ is a super-martingale, i.e., $\mathbb{E}\left[V_\varepsilon(\rho_{k+1}) \mid \rho_k=\rho, u=0\right] - V_\varepsilon(\rho) \leq 0$. Consequently, because $0\in[-\bar{u},\bar{u}]$, we have that $V_\varepsilon(\rho_k)\eval_{u=u_k}$ under the control law \eqref{eq:fblaw_amini2013} is also a super-martingale. 

For any control Hamiltonian $H_1$, we associate the matrix $R$ with elements \citep{amini2013}
\begin{align}
	R_{i,j} = 2\begin{cases}
		\abs{\!\mel{i}{H_1}{j}}^2,& i\neq j \\
		\abs{\!\mel{i}{H_1}{j}}^2 - \mel{i}{H_1^2}{j}, & i=j
	\end{cases},\label{eq:R_def}
\end{align}
which is characterized by having non-negative off-diagonal entries and non-positive diagonal entries. Considering the control Hamiltonian as representing the connectivity of an underlying graph, the matrix $R$ bears resemblance to the Laplacian matrix from graph theory.  

In \cite{amini2011,amini2013}, the authors show that if the vector $\sigma:=\mathrm{diag}(P)$ (i.e., the vector of diagonal elements of $P$) is the solution to the linear systems of equations 
\begin{align}
	R\sigma = \lambda,\label{eq:Rsigma}
\end{align}
for some vector $\lambda$ satisfying
\begin{align}
	\lambda_n < 0\text{ for } n\neq n^\star\quad  \text{and} \quad \lambda_{n^{\star}} = -\sum_{n\neq n^{\star}} \lambda_n > 0,\label{eq:lambda}
\end{align}
then
\begin{align}
	\dod[2]{\mathbb{E}\left[V_\varepsilon(\rho_{k+1}) \mid \rho_k=\op{n}, u\right]}{u}\eval_{u=0} = \lambda_n.
\end{align}
This means that $\mathbb{E}\left[V_\varepsilon(\rho_{k+1}) \mid \rho_k=\op{n}, u\right]$ has a strict maximum at $u=0$ for ${n\neq n^\star}$ and a strict minimum for $n=n^\star$. Consequently, the feedback law \eqref{eq:fblaw_amini2013} will give $u_k\neq 0$ whenever $\rho_k\neq\rho^\star$ and give $u_k=0$ only when $\rho_k=\rho^\star$. This, together with the fact that $V_\varepsilon(\rho)$ is a super-martingale, can be used to prove almost sure convergence to the state $\op{n^\star}$, as presented in \cite{amini2013}.

From \cite[Theorem~2]{amini2013}, we thus have the following proposition.
\begin{proposition}\label{prop:1}
	Consider the Markov process \eqref{eq:qnd_model}. If $P=\mathrm{diag}(\sigma)$ with $\sigma$ being a solution to \eqref{eq:Rsigma} and \eqref{eq:lambda}, then, with the feedback control law \eqref{eq:fblaw_amini2013}, the closed-loop system converges almost surely to $\rho^\star = \argmin_\rho V(\rho)$.
\end{proposition}

\begin{remark}\label{remark:qnd_H0}
	If we consider a model with drift Hamiltonian $H_0$, we have that Proposition \ref{prop:1} still holds, and the $R$ matrix will now be given by $R_{i,j}=2\abs{\mel{i}{H_1 \mathrm{e}^{-iH_0}}{j}} - 2\delta_{i,j} \mel{i}{H_1^2\mathrm{e}^{-iH_0}}{j}$. When constructing the control and drift Hamiltonian from $R$, there will now be considerable freedom in the choice of $H_0$. 
\end{remark}

To summarize, when introducing QND measurements in the quantum system, we have been able to discard Assumptions \ref{assumption:2} and \ref{assumption:3}, but at the cost of having to adhere to Assumption \ref{assumption:qnd} and a slightly more complicated control law. An additional benefit is also that we ensure convergence to a pure state, regardless of the spectrum of $\rho_0$. 

\section{Constructing control Hamiltonians based on $P$}\label{sec:sdp}

We now return to the original problem, namely, \emph{given a fixed matrix $P$ and energy (Lyapunov) function $V(\rho)=\Tr(P\rho)$, can we design a quantum system that ensures convergence to the state which minimizes $V(\rho)$?} So far, we have established sufficient conditions on $P$ and $U(u)$ to ensure convergence to the minimum eigenstate for QND systems. In the following, we will use these conditions to construct a Hamiltonian structure that ensures convergence to the minimum of an arbitrary but known $P$ matrix.

In particular, for $\sigma=\mathrm{diag}(P)$, for which the $n^\star$'th element $\sigma_{n^\star}$ is the minimum, we want to find a matrix $R$ such that $R\sigma=\lambda$ for some $\lambda$ adhering to $\lambda_n < 0$ for $n\neq n^\star$ and $\lambda_{n^{\star}} =-\sum_{n\neq n^{\star}} \lambda_n > 0$. Given the matrix $R$, we can, based on \eqref{eq:R_def}, find a (non-unique) control Hamiltonian $H_1$ with matrix elements
\begin{align}\label{eq:construct_H}
	H_{1}^{i,j} = \begin{cases}
		1/2\sqrt{\!\mel{i}{R}{j}},& i\neq j \\
		0, & i=j
	\end{cases},
\end{align}
where the square root of $R_{i,j}$ may be chosen to be either positive, negative, or imaginary, as long as $H_1$ is Hermitian.

For finding a suitable matrix $R$, we want to minimize the term $\norm{R\sigma - \lambda}$ for some suitable norm. The set of valid $R$ matrices is the set of negative semidefinite matrices with non-negative diagonal elements and non-positive off-diagonal elements and whose column sum (and row sum, by symmetry) is zero, i.e., the set 
\begin{multline}
	\mathcal{R}=\Big\{R\in\mathbb{R}^{n\times n} \mid R\leq 0,\quad \sum_i R_{i,j}= 0,\\ R_{i,i} \leq 0 \quad \forall i, \quad R_{i,j}\geq 0 \quad \forall i < j \Big\},
\end{multline}
which is a convex cone. We thus formulate the following convex optimization problem:
\begin{mini!}|s|
	{R,\lambda}{||R\sigma - \lambda||}
	{\label{eq:SDP1}}{}
	\addConstraint{R\in\mathcal{R}}
	\addConstraint{\lambda_n \leq -\gamma_1, \quad n\neq {n}^\star}
	\addConstraint{\lambda_n \geq \gamma_2, \quad \phantom{-}n={n}^\star },
\end{mini!}
where $\gamma_1,\gamma_2>0$ are hyper-parameters that can be tuned to avoid having $\lambda=0$ as a solution.

For the $\ell_1$-norm, this problem is a standard semidefinite program (SDP), and for the $\ell_2$-norm, this can be shown to be equivalent to a second-order cone program (SOCP) \citep{boyd2004}. In both cases, the problem can be solved using standard tools, e.g., YALMIP \citep{yalmip} with any of the built-in SDP solvers.

We note that the set of $(R,\lambda)$ that satisfies \eqref{eq:Rsigma} is an open set, and the SDP \eqref{eq:SDP1}, for fixed parameters, gives just a single solution. We also note that the solution $(R^\star,\lambda^\star)$ to the SDP may not necessarily give $\norm{R^\star\sigma-\lambda^\star}=0$. One should therefore check that $\tilde{\lambda}:=R^\star \sigma$ satisfies the condition \eqref{eq:lambda}; if this is not satisfied, one could adjust the hyper-parameters in the SDP and test the new solution. The posed SDP may thus, equivalently, be interpreted as a feasibility problem: If \eqref{eq:SDP1} has a feasible solution (in the sense that it satisfies condition \eqref{eq:lambda}), then there exists a system of the form \eqref{eq:qnd_model} with feedback law \eqref{eq:fblaw_amini2013} that can stabilize the state at $\rho^\star = \argmin_\rho \Tr(P\rho)$.

To promote solutions with specific, desirable properties, in particular sparse solutions, we can add different regularizing terms. Here, sparsity in $R$ corresponds to a low number of edges in the underlying graph of the control Hamiltonian. To promote sparse solutions, we minimize the cardinality of $R$ (or the $\ell_0$-pseudo-norm of the vectorized $R$, i.e., $\norm{\vect(R)}_0$). However, including this term in the objective function will result in a non-convex problem. Instead, we will use the convex envelope, namely, $||\mathrm{vec}(R)||_1$ \citep{boyd2004}, resulting in the SDP
\begin{mini!}|s|
	{R,\lambda}{\alpha_1\norm{R\sigma - \lambda} + \alpha_2\norm{\mathrm{vec}(R)}_1}
	{\label{eq:SDP2}}{}
	\addConstraint{R\in\mathcal{R}}
	\addConstraint{\lambda_n \leq -\gamma_1, \quad n\neq {n}^\star}
	\addConstraint{\lambda_n \geq \gamma_2, \quad \phantom{-}n={n}^\star },
\end{mini!}
with $\alpha_i$ being hyper-parameters.

There are many possibilities in forming the SDP to specific needs. For example, to control the scaling of the numerical values, a constraint like $\Tr(R)\leq \beta$ for $\beta>0$ could be added. One could also include different constraints on $R$ to enforce particular structures, e.g., constraining individual elements to be zero to get a structure that lends itself to a decomposition in terms of elementary qubit gates. We will, however, not investigate these possibilities further here.


To summarize, the full synthesis procedure is as follows:
\begin{enumerate}
	\item Given diagonal matrix $P$ with $\sigma$ the vector of diagonal elements, solve the SDP \eqref{eq:SDP2} to find $R^\star$.
	\item Check that $\tilde{\lambda} = R^\star\sigma$ satisfies $\tilde{\lambda}_n < 0$ for $n\neq n^\star$ and $\tilde{\lambda}_{n^{\star}} = -\sum_{n\neq n^{\star}} \tilde{\lambda}_n > 0$. If not, return to point 1 and adjust the hyper-parameters.
	\item Construct a (Hermitian) control Hamiltonian $H_1^\star$ using \eqref{eq:construct_H}.
	\item Using the feedback law \eqref{eq:fblaw_amini2013}, the state $\rho_k$ of system \eqref{eq:qnd_model} almost surely converges to $\rho^\star = \argmin_\rho V(\rho)$.
\end{enumerate}

\section{A numerical example}\label{sec:example}
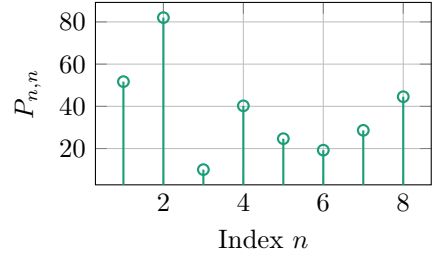
\begin{figure}[ht]
	\centering
	\tikzsetnextfilename{P}
	\begin{tikzpicture}[trim axis left, trim axis right]
		\begin{axis}[xmajorgrids,ymajorgrids,height=4cm,width=6cm,ylabel={$P_{n,n}$}, xlabel={Index $n$}]
			\addplot+[ycomb,mark=o] table[] {
				1.0000   51.7022
				2.0000   82.0324
				3.0000   10.0114
				4.0000   40.2333
				5.0000   24.6756
				6.0000   19.2339
				7.0000   28.6260
				8.0000   44.5561
			};
		\end{axis}
	\end{tikzpicture}
	\caption{The diagonal elements of $P$ used in the example.}
	\label{fig:P_elements}
\end{figure}

In this section, we provide a numerical example showcasing the methodology from Section \ref{sec:sdp}. We consider a system with dimension $N=8$ and with the $P$ matrix with diagonal elements given as shown in Figure \ref{fig:P_elements}. As can be seen, the minimum is at $n^\star=3$.

We solve the SDP \eqref{eq:SDP1} in \textsc{Matlab} using \textsc{Yalmip} \citep{yalmip} with \textsc{Mosek} \citep{mosek} as SDP solver. In the following, we show the results for two cases: The solution to problem \eqref{eq:SDP1} with $\gamma=1$ and the solution to the sparsity-promoting problem \eqref{eq:SDP2} with $\gamma=1,\alpha_1=\alpha_2=1$. We observed no noticeable difference in the structure of $R$ when comparing solutions to the problems with the $\ell_1$-norm and $\ell_2$-norm, although the numerical values were slightly different. The results shown here have been obtained using the $\ell_2$-norm.

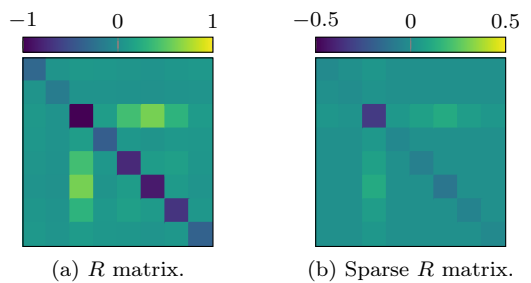
\begin{figure}[ht]
	\centering
	\subfloat[$R$ matrix.]{\tikzsetnextfilename{R_1}
\begin{tikzpicture}[trim axis left, trim axis right]
	\begin{axis}[enlargelimits=false,colorbar horizontal,colormap/viridis,point meta min=-1, point meta max=1,xmax=8.5, xmin=0.5, ymax=8.5, ymin=0.5,scale only axis, height=2.5cm, width=2.5cm, ytick=0,xtick=0,
	colorbar style={
			at={(0.5,1.03)},
			anchor=south,
			xticklabel pos=upper,
			xtick={-1,0,1},
			height=0.2cm,
			xticklabel style={font=\small},
		}]
		\addplot [
		matrix plot,
		point meta=explicit,
		mesh/cols=8,
		] coordinates {
			(1,1) [-0.33484] (1,2) [0.03407] (1,3) [0.068106] (1,4) [0.056603] (1,5) [0.039627] (1,6) [0.021134] (1,7) [0.052815] (1,8) [0.062481] 
			(2,1) [0.03407] (2,2) [-0.16678] (2,3) [0.028492] (2,4) [0.026086] (2,5) [0.016806] (2,6) [0.009088] (2,7) [0.022617] (2,8) [0.029618] 
			(3,1) [0.068106] (3,2) [0.028492] (3,3) [-1.5646] (3,4) [0.10791] (3,5) [0.39185] (3,6) [0.58714] (3,7) [0.29354] (3,8) [0.087559] 
			(4,1) [0.056603] (4,2) [0.026086] (4,3) [0.10791] (4,4) [-0.40668] (4,5) [0.055694] (4,6) [0.03501] (4,7) [0.066148] (4,8) [0.059236] 
			(5,1) [0.039627] (5,2) [0.016806] (5,3) [0.39185] (5,4) [0.055694] (5,5) [-0.77519] (5,6) [0.09859] (5,7) [0.12359] (5,8) [0.049032] 
			(6,1) [0.021134] (6,2) [0.009088] (6,3) [0.58714] (6,4) [0.03501] (6,5) [0.09859] (6,6) [-0.86139] (6,7) [0.081741] (6,8) [0.02869] 
			(7,1) [0.052815] (7,2) [0.022617] (7,3) [0.29354] (7,4) [0.066148] (7,5) [0.12359] (7,6) [0.081741] (7,7) [-0.70175] (7,8) [0.061296] 
			(8,1) [0.062481] (8,2) [0.029618] (8,3) [0.087559] (8,4) [0.059236] (8,5) [0.049032] (8,6) [0.02869] (8,7) [0.061296] (8,8) [-0.37791] 
		};
	\end{axis}
\end{tikzpicture}} \hspace{1cm}
	\subfloat[Sparse $R$ matrix.]{\tikzsetnextfilename{R_sparse_1}
\begin{tikzpicture}[trim axis left, trim axis right]
		\begin{axis}[enlargelimits=false,colorbar horizontal,colormap/viridis,point meta min=-.5, point meta max=0.5,xmax=8.5, xmin=0.5, ymax=8.5, ymin=0.5,scale only axis, height=2.5cm, width=2.5cm, ytick=0,xtick=0,
		colorbar style={
			at={(0.5,1.03)},
			anchor=south,
			xticklabel pos=upper,
			xtick={-.5,0,.5},
			height=0.2cm,
			xticklabel style={font=\small},
		}]
		\addplot [
		matrix plot,
		point meta=explicit,
		mesh/cols=8,
		] coordinates {
			(1,1) [-0.023986] (1,2) [1.8645e-12] (1,3) [0.023986] (1,4) [2.2714e-12] (1,5) [7.3437e-14] (1,6) [-4.9203e-14] (1,7) [1.5454e-13] (1,8) [2.3435e-12] 
			(2,1) [1.8645e-12] (2,2) [-0.013885] (2,3) [0.013885] (2,4) [6.4492e-13] (2,5) [1.794e-14] (2,6) [-2.4499e-14] (2,7) [3.7073e-14] (2,8) [9.1215e-13] 
			(3,1) [0.023986] (3,2) [0.013885] (3,3) [-0.33025] (3,4) [0.033089] (3,5) [0.068193] (3,6) [0.10843] (3,7) [0.053721] (3,8) [0.028948] 
			(4,1) [2.2714e-12] (4,2) [6.4492e-13] (4,3) [0.033089] (4,4) [-0.033089] (4,5) [3.0032e-13] (4,6) [8.0501e-14] (4,7) [5.433e-13] (4,8) [3.1296e-12] 
			(5,1) [7.3437e-14] (5,2) [1.794e-14] (5,3) [0.068193] (5,4) [3.0032e-13] (5,5) [-0.068193] (5,6) [4.7222e-13] (5,7) [4.4531e-13] (5,8) [1.8792e-13] 
			(6,1) [-4.9203e-14] (6,2) [-2.4499e-14] (6,3) [0.10843] (6,4) [8.0501e-14] (6,5) [4.7222e-13] (6,6) [-0.10843] (6,7) [2.1741e-13] (6,8) [2.1632e-14] 
			(7,1) [1.5454e-13] (7,2) [3.7073e-14] (7,3) [0.053721] (7,4) [5.433e-13] (7,5) [4.4531e-13] (7,6) [2.1741e-13] (7,7) [-0.053721] (7,8) [3.2333e-13] 
			(8,1) [2.3435e-12] (8,2) [9.1215e-13] (8,3) [0.028948] (8,4) [3.1296e-12] (8,5) [1.8792e-13] (8,6) [2.1632e-14] (8,7) [3.2333e-13] (8,8) [-0.028948] 
		};
	\end{axis}
\end{tikzpicture}}
	\caption{The elements of the $R$ matrix resulting from the solution to the SDP from the example.}
	\label{fig:R}
\end{figure}

\begin{figure}[ht]
	\centering
	\tikzsetnextfilename{lambda}
	\begin{tikzpicture}[trim axis left, trim axis right]
		\begin{axis}[xmajorgrids, ymajorgrids,ymax=30,ymin=-10,height=4cm,width=6cm, xlabel={Index $n$}, ylabel={$\lambda_n$}]
			\addplot+[ycomb,mark=o] table[] {
				1.0000   -5.8776
				2.0000   -8.0283
				3.0000   27.8025
				4.0000   -3.6350
				5.0000   -1.9182
				6.0000   -1.3921
				7.0000   -2.5492
				8.0000   -4.4021
			};
			\addplot+[ycomb,mark=o] table[] {
				1.1000   -1
				2.1000   -1
				3.1000   7
				4.1000   -1
				5.1000   -1
				6.1000   -1
				7.1000   -1
				8.1000   -1
			};
			\legend{Non-sparse,Sparse}
		\end{axis}
	\end{tikzpicture}
	\caption{The elements of vector $\lambda$ resulting from the numerical example.}
	\label{fig:lambda}
\end{figure}
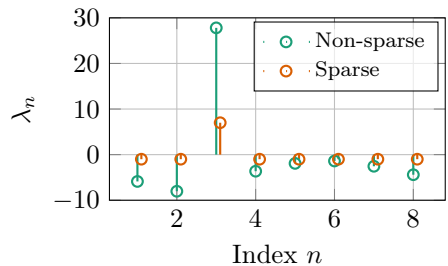

In Figure \ref{fig:R}, the resulting $R$ matrix is shown for both the non-sparse and sparse solution. Additionally, the resulting values of $\lambda$ are shown in Figure \ref{fig:lambda}. We note that the objective function at our solution is $\norm{R^\star\sigma - \lambda^\star}_2\approx 0$ in both cases.

As can be seen, the resulting matrix differs in the two cases, where it is clear to see that the sparsity-promoting SDP results in a sparse solution.

\begin{figure}[ht]
	\centering
	\subfloat[$H_1$ matrix.]{\tikzsetnextfilename{H1_1}
\begin{tikzpicture}[trim axis left, trim axis right]
		\begin{axis}[enlargelimits=false,colorbar horizontal,colormap/viridis,point meta min=-0.5, point meta max=0.5,xmax=8.5, xmin=0.5, ymax=8.5, ymin=0.5,scale only axis, height=2.5cm, width=2.5cm, ytick=0,xtick=0,
		colorbar style={
			at={(0.5,1.03)},
			anchor=south,
			xticklabel pos=upper,
			xtick={-0.5,0,0.5},
			height=0.2cm,
			xticklabel style={font=\small},
		}]
		\addplot [
		matrix plot,
		point meta=explicit,
		mesh/cols=8,
		] coordinates {
			(1,1) [0] (1,2) [0.13052] (1,3) [0.18453] (1,4) [0.16823] (1,5) [0.14076] (1,6) [0.10279] (1,7) [0.1625] (1,8) [0.17675] 
			(2,1) [0.13052] (2,2) [0] (2,3) [0.11936] (2,4) [0.11421] (2,5) [0.091668] (2,6) [0.067409] (2,7) [0.10634] (2,8) [0.12169] 
			(3,1) [0.18453] (3,2) [0.11936] (3,3) [0] (3,4) [0.23228] (3,5) [0.44263] (3,6) [0.54182] (3,7) [0.38311] (3,8) [0.20924] 
			(4,1) [0.16823] (4,2) [0.11421] (4,3) [0.23228] (4,4) [0] (4,5) [0.16687] (4,6) [0.13231] (4,7) [0.18186] (4,8) [0.1721] 
			(5,1) [0.14076] (5,2) [0.091668] (5,3) [0.44263] (5,4) [0.16687] (5,5) [0] (5,6) [0.22203] (5,7) [0.24859] (5,8) [0.15658] 
			(6,1) [0.10279] (6,2) [0.067409] (6,3) [0.54182] (6,4) [0.13231] (6,5) [0.22203] (6,6) [0] (6,7) [0.20217] (6,8) [0.11977] 
			(7,1) [0.1625] (7,2) [0.10634] (7,3) [0.38311] (7,4) [0.18186] (7,5) [0.24859] (7,6) [0.20217] (7,7) [0] (7,8) [0.17507] 
			(8,1) [0.17675] (8,2) [0.12169] (8,3) [0.20924] (8,4) [0.1721] (8,5) [0.15658] (8,6) [0.11977] (8,7) [0.17507] (8,8) [0] 
		};
	\end{axis}
\end{tikzpicture}} \hspace{1cm}
	\subfloat[Sparse $H_1$matrix.]{\tikzsetnextfilename{H1_sparse_1}
\begin{tikzpicture}[trim axis left, trim axis right]
		\begin{axis}[enlargelimits=false,colorbar horizontal,colormap/viridis,point meta min=-0.5, point meta max=0.5,xmax=8.5, xmin=0.5, ymax=8.5, ymin=0.5,scale only axis, height=2.5cm, width=2.5cm, ytick=0,xtick=0,
		colorbar style={
			at={(0.5,1.03)},
			anchor=south,
			xticklabel pos=upper,
			xtick={-.5,0,.5},
			height=0.2cm,
			xticklabel style={font=\small},
		}]
		\addplot [
		matrix plot,
		point meta=explicit,
		mesh/cols=8,
		] coordinates {
			(1,1) [0] (1,2) [9.6554e-07] (1,3) [0.10951] (1,4) [1.0657e-06] (1,5) [1.9162e-07] (1,6) [1.5685e-07] (1,7) [2.7797e-07] (1,8) [1.0825e-06] 
			(2,1) [9.6554e-07] (2,2) [0] (2,3) [0.083321] (2,4) [5.6786e-07] (2,5) [9.4709e-08] (2,6) [1.1068e-07] (2,7) [1.3615e-07] (2,8) [6.7533e-07] 
			(3,1) [0.10951] (3,2) [0.083321] (3,3) [0] (3,4) [0.12862] (3,5) [0.18465] (3,6) [0.23284] (3,7) [0.16389] (3,8) [0.12031] 
			(4,1) [1.0657e-06] (4,2) [5.6786e-07] (4,3) [0.12862] (4,4) [0] (4,5) [3.875e-07] (4,6) [2.0063e-07] (4,7) [5.212e-07] (4,8) [1.2509e-06] 
			(5,1) [1.9162e-07] (5,2) [9.4709e-08] (5,3) [0.18465] (5,4) [3.875e-07] (5,5) [0] (5,6) [4.8591e-07] (5,7) [4.7187e-07] (5,8) [3.0653e-07] 
			(6,1) [1.5685e-07] (6,2) [1.1068e-07] (6,3) [0.23284] (6,4) [2.0063e-07] (6,5) [4.8591e-07] (6,6) [0] (6,7) [3.2971e-07] (6,8) [1.04e-07] 
			(7,1) [2.7797e-07] (7,2) [1.3615e-07] (7,3) [0.16389] (7,4) [5.212e-07] (7,5) [4.7187e-07] (7,6) [3.2971e-07] (7,7) [0] (7,8) [4.0208e-07] 
			(8,1) [1.0825e-06] (8,2) [6.7533e-07] (8,3) [0.12031] (8,4) [1.2509e-06] (8,5) [3.0653e-07] (8,6) [1.04e-07] (8,7) [4.0208e-07] (8,8) [0] 
		};
	\end{axis}
\end{tikzpicture}}
	\caption{The elements of control Hamiltonain $H_1$ constructed in the example.}
	\label{fig:H1}
\end{figure}
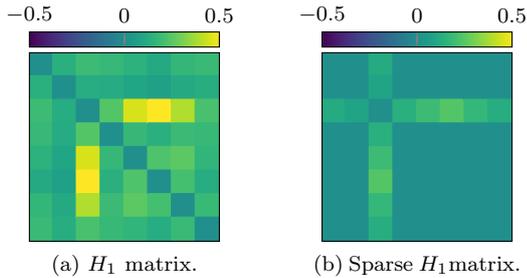

In Figure \ref{fig:H1}, the resulting control Hamiltonian $H_1$ is shown, where we have chosen all entries to be real. In terms of connections between basis states, it can be seen that the sparse solution only has connections to the basis state that minimizes $P$, i.e., $n^\star=3$. The control Hamiltonian thus clearly does not satisfy Assumption \ref{assumption:3} from Section \ref{sec:deterministic}, but, as previously stated, this was never a requirement in the QND system.

We also note that the the numerical values in both cases are in the interval $[-1,\,1]$. To promote solutions with numerically larger values, it is possible to adjust the optimization problem in terms of the parameter $\gamma$ or by adding additional constraints, e.g., an inequality constraint on the trace such as $\Tr(R)\leq \beta$ for $\beta < 0$.

For the QND measurements, we use a parameterization similar to that of the photon box experiment \citep{amini2012}, namely,
\begin{equation}
\begin{aligned}
	M_0 = \sum_n \cos(\phi_0 + n\theta) \op{n},\\ M_1 = \sum_n \sin(\phi_0 + n\theta) \op{n},
\end{aligned}
\end{equation}
where we pick $\phi_0=\frac{1}{8}$ and $\theta=\frac{\pi}{4}$.

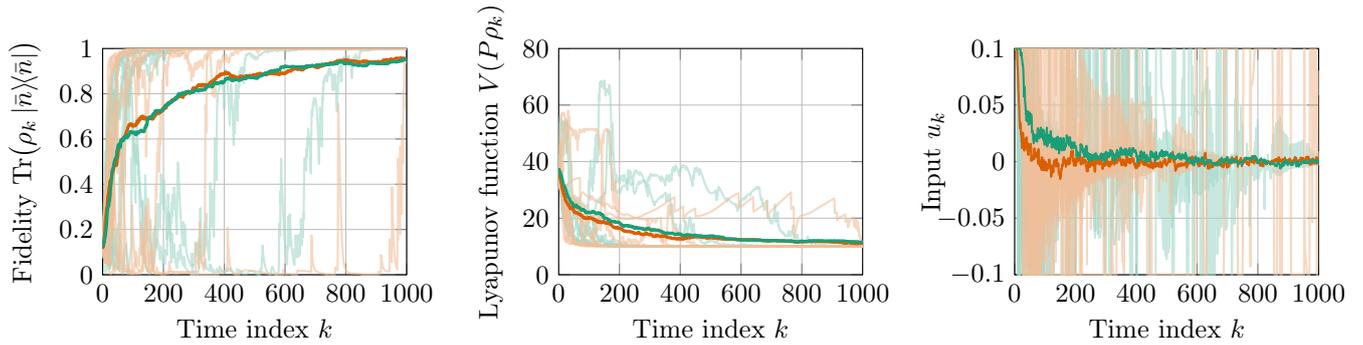
\begin{figure*}[ht]
	\centering
	\tikzsetnextfilename{qnd_traj_sparse_and_non-sparse_mixed_fmincon}
\begin{tikzpicture}
	\begin{groupplot}[group style={group size=3 by 1, horizontal sep=2cm, vertical sep=1.5cm, xlabels at=edge bottom},height=3cm,width=4cm, ylabel near ticks, xlabel near ticks, xmajorgrids, ymajorgrids, scale only axis, enlargelimits=false,, enlarge y limits=false, legend pos=north east,xlabel={Time index $k$}, xmin=0, ymin=0, xmax=1000]
		
		\nextgroupplot[ymax=1, legend pos=outer north east,ylabel={Fidelity $\Tr(\rho_k \op{\bar{n}})$}]
		\foreach \column in {10,...,19}{
			\addplot+[.!40!white,opacity=0.6,forget plot] table[x index=0,y index={\column}] {../../../Matlab/plotdata/qnd_traj_1_fmincon/F3.dat};
		}
		\pgfplotsset{cycle list shift=1}
		\foreach \column in {10,...,19}{
			\addplot+[.!40!white,opacity=0.6, forget plot] table[x index=0,y index={\column}] {../../../Matlab/plotdata/qnd_traj_sparse_1_fmincon/F3.dat};
		}
		\pgfplotsset{cycle list shift=-1}
		\addplot+[very thick] file {../../../Matlab/plotdata/qnd_traj_1_fmincon/F3_mean.dat};
		\addplot+[very thick] file {../../../Matlab/plotdata/qnd_traj_sparse_1_fmincon/F3_mean.dat};
		
		\nextgroupplot[legend pos=outer north east, ylabel={Lyapunov function $V(P\rho_k)$}, ymax=80]
		\foreach \column in {10,...,19}{
			\addplot+[.!40!white,opacity=0.6,forget plot] table[x index=0,y index={\column}] {../../../Matlab/plotdata/qnd_traj_1_fmincon/V.dat};
		}
		\pgfplotsset{cycle list shift=1}
		\foreach \column in {10,...,19}{
			\addplot+[.!40!white,opacity=0.6,forget plot] table[x index=0,y index={\column}] {../../../Matlab/plotdata/qnd_traj_sparse_1_fmincon/V.dat};
		}
		\pgfplotsset{cycle list shift=-1}
		\addplot+[very thick] file {../../../Matlab/plotdata/qnd_traj_1_fmincon/V_mean.dat};
		\addplot+[very thick] file {../../../Matlab/plotdata/qnd_traj_sparse_1_fmincon/V_mean.dat};
		
		\nextgroupplot[xlabel=Time index $k$, legend pos=outer north east,ymin=-0.1,ymax=0.1, ylabel=Input $u_k$, ylabel shift=-1em]
		\foreach \column in {10,...,19}{
			\addplot+[.!40!white,opacity=0.6,forget plot] table[x index=0,y index={\column}] {../../../Matlab/plotdata/qnd_traj_1_fmincon/u.dat};
		}
		\pgfplotsset{cycle list shift=1}
		\foreach \column in {10,...,19}{
			\addplot+[.!40!white, opacity=0.6,forget plot] table[x index=0,y index={\column}] {../../../Matlab/plotdata/qnd_traj_sparse_1_fmincon/u.dat};
		}
		\pgfplotsset{cycle list shift=-1}
		\addplot+[] file {../../../Matlab/plotdata/qnd_traj_1_fmincon/u_mean.dat};
		\addplot+[] file {../../../Matlab/plotdata/qnd_traj_sparse_1_fmincon/u_mean.dat};
		
	\end{groupplot}
\end{tikzpicture}%
	\caption{The mean trajectory (dark-colored) of 100 realizations, and trajectories of 10 realizations (light-colored), shown both for the non-sparse (\textcolor{Dark2-A}{\rule[.5ex]{1em}{2pt}}) and sparse (\textcolor{Dark2-B}{\rule[.5ex]{1em}{2pt}}) control Hamiltonians.}
	\label{fig:state1}
\end{figure*}

In Figure \ref{fig:state1}, the mean trajectory of 100 realizations as well as sample trajectories of 10 realizations are shown with the resulting control Hamiltonians. All realizations are initialized with $\rho_0=\frac{1}{2}\op{0} + \frac{1}{2N} \left(\sum_i \left(\ket{i}\right)\right) \left(\sum_i \left(\bra{i}\right)\right)$, i.e., a mixed state with probability $1/2$ of being in $\op{0}$ and probability $1/2$ in uniform superposition over all basis states.

In the feedback law \eqref{eq:fblaw_amini2013}, we observed no noticeable difference between having $\varepsilon=0$ or $\varepsilon$ being a small number, which agrees with the observations made in \cite{amini2013}. Thus, in the simulations shown here we have $\varepsilon=0$. 

Likewise, as in \cite{amini2013}, we use a quadratic approximation of the nonlinear optimization problem in the feedback law \eqref{eq:fblaw_amini2013} to speed up computations. In particular, we solve the quadratic program
\begin{multline}
	u_k = \argmin_{u\in[-\bar{u},\,\bar{u}]} \frac{1}{2} \bigg(\Tr(\left[[H_1,P],H_1\right]\rho_k)\\ -\frac{\varepsilon}{4} \sum_i \left(\!\ev{[H_1,\rho_k]}{i}\right)^2\bigg) u^2 -i\Tr([H_1,P]\rho_k)u\label{eq:QP}
\end{multline}
with $\bar{u}=0.1$. 

As can be seen from Figure \ref{fig:state1}, the state converges to the state $\rho^\star=\op{\bar{n}}$ in almost all cases within the first 1000 iterations. Comparing the non-sparse and sparse solutions, it can be seen that, on average, they are almost identical.



\section{Discussion and future work}\label{sec:discussion}
Relating our results to VQAs, it would be necessary to adhere to the constraints of the quantum hardware. For instance, given a control Hamiltonian $H_1$, can we efficiently describe the resulting unitary operation in terms of elementary quantum (qubit) gates? This is already a well-established research area within quantum simulation (see, e.g., Section 4.7 in \cite{nielsen} for an introduction). In particular, if the Hamiltonian is sparse in the Pauli basis, it can be efficiently represented using elementary qubit operations.

Likewise, QND measurements can in general be realized by entangling an ancillary qubit to the principal quantum circuit in a proper way, and then perform a projective measurement on the ancillary qubit. However, such an operation would have to be repeated in the circuit for every time instance $k$. Alternatively, one could investigate hardware solutions, where QND measurements are cheap, e.g., in optical quantum computing \citep{munro2005}.

\vspace*{-0.15em}
In the present work, we have assumed the state $\rho_k$ to be known perfectly at every time instance $k$ in order for us to compute the feedback control. In practice, this is not a valid assumption. However, using the known QND measurement outcomes, it is possible to estimate the state at time $k$ using an observer / quantum filter. As shown in \cite{amini2013}, such observers are robust to uncertainties in the initialization of the filter, and the state feedback based on the state estimate from the filter will ensure convergence under a very mild assumption on the initial state estimate. Although not shown here, we observed similar convergence in our numerical example when basing the state feedback on such a state estimate.

\vspace*{-0.15em}
Another option is to adopt an approach similar to that of \cite{magann2022}, where the necessary quantity in the feedback law is estimated based on several experiments of each time step. In particular, if we consider the coefficients from the quadratic optimization problem \eqref{eq:QP}, i.e., $\Tr([H_1,P]\rho_k)$ and $\Tr(\left[[H_1,P],H_1\right]\rho_k)$, as the expectation values of the observables $[H_1,P]$ and $\left[[H_1,P],H_1\right]$, respectively, we can compute a sample estimate of these by measuring these observables many times. This, however, requires the time evolution from $k$ to $k+1$ to be repeated many times, as the state is collapsed after each measurement of the observable. Note also that this assumes that the aforementioned observables can be expressed in terms of easy-to-measure observables on the quantum computer, e.g., in terms of Pauli gates. The posed SDP can easily be adjusted to accommodate such structures by including additional constraints on $R$, which, as long as they a linear, will preserve the convexity of the problem.

\vspace*{-0.15em}
In our work, we also observed that measurement-free systems with drift (i.e., non-zero $H_0$), as presented in Section \ref{sec:deterministic}, also converge to the minimum eigenstate using the methodology presented in Section \ref{sec:sdp}. In particular, we conjecture that Assumption \ref{assumption:3} can be relaxed in this case, which might be of particular interest to the aforementioned approach by \cite{magann2022}. In future work, we will investigate this in a more rigorous manner.


\vspace*{-0.15em}
Finally, it should be noted that the methodology as presented here will not be directly applicable in VQA situations in its current form. The motivation for using, e.g., the QAOA in the first place is the intractability of using classical computers to solve the problem of finding the minimum eigenvalue. Here, we use the $P$ matrix explicitly in our synthesis of $R$, which would be intractable for problems of any relevant size. In future work, this will have to be addressed. However, in its current form, one can use the presented methodology to perform analyses of small-scale instances of specific VQA problem formulations and investigate whether (part of) the structure is the same regardless of the problem size (see, e.g., \cite{streif2020} for results suggesting such relations).


%
%
%
%


%

\bibliography{../../bibtex/literature.bib} 	
\end{document}